\documentclass[a4paper,fleqn]{cas-sc}

\usepackage[authoryear,longnamesfirst]{natbib}
\usepackage{xcolor} 
\usepackage{subcaption}

\usepackage{lineno}

\def\tsc#1{\csdef{#1}{\textsc{\lowercase{#1}}\xspace}}
\tsc{WGM}
\tsc{QE}

\begin{document}
\let\WriteBookmarks\relax
\def\floatpagepagefraction{1}
\def\textpagefraction{.001}

\shorttitle{Lunar ejecta flux on Earth}    

\shortauthors{Castro-Cisneros, Malhotra, Rosengren}  

\title [mode = title]{Lunar impact ejecta flux on the Earth}  



%

\author[1]{Jose Daniel Castro-Cisneros}[
    orcid=0000-0002-6624-4214
]

\cormark[1]


\ead{jdcastrocisneros@arizona.edu}


\credit{Conceptualization, Data curation, Formal analysis, Investigation, Software, Visualization, Writing - original draft, Writing - review \& editing}

\affiliation[1]{organization={Physics Department, The University of Arizona},
            addressline={1118 E. Fourth Street}, 
            city={Tucson},
            postcode={85721}, 
            state={AZ},
            country={USA}}

\author[2]{Renu Malhotra}




\credit{Conceptualization, Funding acquisition, Supervision, Writing - original draft, Writing - review \& editing}

\affiliation[inst2]{organization={Lunar and Planetary Laboratory, The University of Arizona},
            addressline={1629 E. University Blvd.}, 
            city={Tucson},
            postcode={85721}, 
            state={AZ},
            country={USA}}


\author[3]{Aaron J. Rosengren}
\credit{Supervision, Writing - original draft, Writing - review \& editing}

\affiliation[inst3]{organization={Mechanical and Aerospace Engineering, UC San Diego},
            addressline={9500 Gilman Drive}, 
            city={La Jolla},
            postcode={92093}, 
            state={CA},
            country={USA}}

\cortext[cor1]{Corresponding author}


\begin{abstract}
The transfer of material between planetary bodies due to impact events is important for understanding planetary evolution, meteoroid impact fluxes, the formation of near-Earth objects (NEOs), and even the provenance of volatile and organic materials at Earth. This study investigates the dynamics and fate of lunar ejecta  reaching Earth. 
We employ the high-accuracy IAS15 integrator within the REBOUND package to track for 100,000 years the trajectories of 6,000 test particles launched from various lunar latitudes and longitudes. 
Our model incorporates a realistic velocity distribution for ejecta fragments (tens of meters in size), derived from large lunar cratering events. 
Our results show that 22.6\% of lunar ejecta collide with Earth, following a power-law $C(t) \propto   t^{0.315}$, with half of the impacts occurring within $\sim$10,000 years. 
We also confirm that impact events on the Moon’s trailing hemisphere serve as a dominant source of Earth-bound ejecta, consistent with previous studies. 
Additionally, a small fraction of ejecta remains transiently in near-Earth space, providing evidence that lunar ejecta may contribute to the NEO population. 
This aligns with recent discoveries of Earth co-orbitals such as Kamo\'oalewa (469219, 2016 HO3) and 2024 PT5, both exhibiting spectral properties consistent with lunar material. 
These findings enhance our understanding of the lunar ejecta flux to Earth, providing insights into the spatial and temporal patterns of this flux and its broader 
influence on 
the near-Earth environment.

\nocite{*}

\end{abstract}


\begin{highlights}

\item Lunar impact ejecta flux to Earth rapidly halves in a timescale of $\sim 10$ kyr.
\item Trailing hemisphere ejecta contribute $\sim 2 \times$  more than leading hemisphere.  
\item Lunar ejecta cumulative Earth collisions follow a power-law over time. 
\item Impact locations cluster near equator; polar-equatorial flux ratio 0.76. 
\item PM/AM impactors are mostly symmetrically distributed, with peaks near 6 PM/AM.

\end{highlights}


\maketitle

\shortcites{bG95,Gladman1996,Castro2023,Castro2025,Jiao2024,bS21,Kareta2025,Popova2013,Collins2005,Gallant2009,Mazrouei2019,Nesvorny2024,Bills1999,Collins2017,Nishiizumi2009,Lorenzetti2005,Joy2023}

\section{Introduction}
\label{sec:Introduction}

The study of impact ejecta plays a fundamental role in understanding the dynamical exchange of material between planetary bodies. 
Since the early works of \cite{bG95,Gladman1996} which investigated the transfer of impact-generated debris among terrestrial planets, there has been increasing interest in the fate of ejected material, particularly from the Moon. 
The present work investigates the launch conditions, transit times, and flux of lunar ejecta reaching Earth.
We utilize updated numerical methods that improve on previous approaches. 
By employing modern numerical integrators and a physically motivated ejection-velocity distribution, we refine earlier estimates of lunar ejecta dynamics and transfer. 
This study extends the work of \cite{Castro2023,Castro2025}, which focused on the evolution of lunar ejecta that enter Earth's co-orbital zones.
We broadened the analysis to include a physically realistic set of initial conditions and a longer integration timescale. 
By systematically investigating the fate of lunar ejecta using advanced computational techniques and a realistic range of launch conditions, this work aims to clarify the time-evolution of the flux of lunar material on Earth and also to evaluate its potential contribution to the population of near-Earth objects (NEOs).\\ 

One of the fundamental challenges in understanding Earth’s impact history is the lack of a direct record of its earliest impact events. In contrast, the lunar surface serves as a well-preserved archive of impact history, offering insights into the impact flux within the Earth-Moon system and 
its implications for planetary evolution \citep{Zahnle2006}. 
In this context, investigating the fate of lunar ejecta aids in reconstructing Earth's impact history and assessing its potential influence on both biological and geological evolution. \\

As a further motivation, the possibility of a population of lunar ejecta orbiting Earth has gained significant interest in recent years. 
One of the most compelling cases is Kamo'oalewa (469219, 2016 HO3), a small NEO approximately 36--100 meters in diameter \citep{Jiao2024}, that has been identified as a quasi-satellite of Earth. 
\cite{bS21} presented observational evidence suggesting that Kamo'oalewa's reflectance spectrum closely matches that of lunar materials, making it a strong candidate for a lunar ejecta fragment. 
This finding was further supported by orbital evolution and hydrodynamic impact simulations  studies conducted by \cite{Castro2023} and \cite{Jiao2024}, respectively, which demonstrated that impact events on the Moon can generate ejecta that temporarily become co-orbital with Earth's heliocentric orbit. 
Their results reinforced the feasibility of a long-lived, lunar-derived population in Earth's vicinity. 
A more recent discovery, 2024 PT5, a 10--12 meter-diameter object initially referred to as a mini-moon \citep{delaFuenteMarcos:2024}, has further strengthened the case for lunar ejecta in near-Earth space. 
\cite{Kareta2025} conducted a spectral analysis of 2024 PT5 and found that its spectral properties closely resemble known lunar samples, making it another strong candidate for a Moon-derived fragment. 
Such objects offer opportunities to study ancient lunar material ejected during past impact events, with the potential to obtain more direct constraints on the hydrodynamics of large impact processes on the Moon.\\ 

The remainder of the paper is structured as follows. Section 2 details the numerical model used in our simulations, including the integrator, initial conditions, and initial velocity distribution. 
Section 3 presents the results, focusing on the flux of lunar ejecta at Earth, their collision times and Earth-impact velocities and their spatial distribution at the time of impact. 
Section 4 discusses the implications of our findings, including the potential biological and geological effects and role of lunar ejecta in the NEO population; 
suggestions of directions for future research are also provided. 
Finally, Section 5 summarizes our conclusions. 

\section{Methodology}
\label{sec:Methodology}

\subsection{Numerical model}
\label{subsec:model}

Building on the numerical methods developed in \cite{Castro2023}, 
we investigate the dynamical evolution and collisional fates of lunar ejecta, 
inspired by the foundational work of \cite{bG95,Gladman1996}.
With the advances in computer technology and in numerical integrators since the time of the previous studies, we can now make use of a higher order, adaptive step-size integrator together with a higher-fidelity physical model for more realistic simulations. 
For example, \cite{bG95} divided the evolution of lunar ejecta into two separate phases: a geocentric phase (including only the Sun, Earth, and Moon) and a heliocentric phase (including the Sun, Earth, and other planets, but excluding the Moon). 
Our approach incorporates all planets and the Moon at all times throughout the simulation. 
We used the \texttt{IAS15} integrator from the \texttt{REBOUND} N-body simulation package \citep{Rein2014}, chosen for its adaptive time-stepping and accurate handling of close encounters.
Another key difference lies in the initial velocity distribution of the ejected particles. 
\cite{bG95} assumed a rather uniform and relatively narrow velocity distribution, whereas we adopt a more physically motivated distribution based on hydrodynamic-impact simulations of a vertical collision. 
By implementing these improvements, our study extends and refines the framework established by \cite{bG95}, while using current state-of-the-art computational tools and a more complete dynamical model. 

Test particles (TPs) were launched from the lunar surface and tracked under the gravitational influence of the Sun, the eight major planets, and the Moon. 
To account for collisions, the \texttt{REBOUND} package’s collision detection capabilities were used with the \texttt{DIRECT} module. 
We adopted an initial time step of 1.2 days, with an accuracy parameter of $\epsilon = 10^{-9}$ (default value). 
Each test particle was propagated for 100,000 years, with orbital elements and Cartesian coordinates recorded every five years for further analysis, as well as at the moment of collision, if one occurred.
A collision is defined as occurring at the top of the atmosphere, set at 100 km above the Earth's equatorial radius of 6,378 km. This choice aligns approximately with the Kármán line, commonly considered the boundary between Earth's atmosphere and outer space \citep{Mcdowell2018}.

\subsection{Initial conditions}
\label{subsec:launching}

We selected three lunar latitudes as launch sites: $70^{\circ}$ North, the Equator, and $53^{\circ}$ South (near the center of the South Pole–Aitken Basin).

At each latitude, we established four launch sites positioned on the Moon's near side, trailing side, far side, and leading side. This arrangement ensures comprehensive representation of the spatial coverage of launch conditions. \\

\begin{figure}[ht]
    \centering
    \begin{subfigure}[b]{0.45\linewidth}
        \centering
        \includegraphics[width=0.8\textwidth]{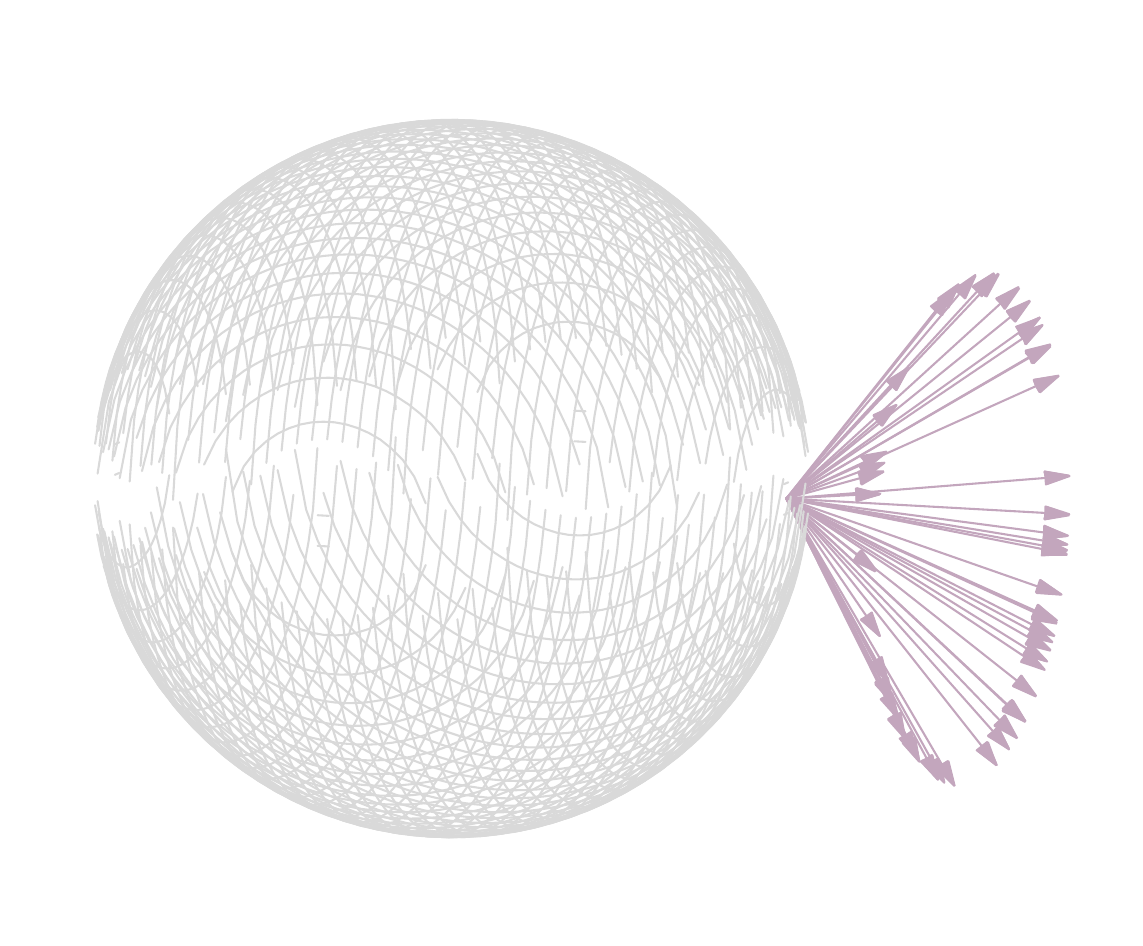}   
    \end{subfigure}
    \begin{subfigure}[b]{0.45\linewidth}
        \centering
        \includegraphics[width=0.6\textwidth]{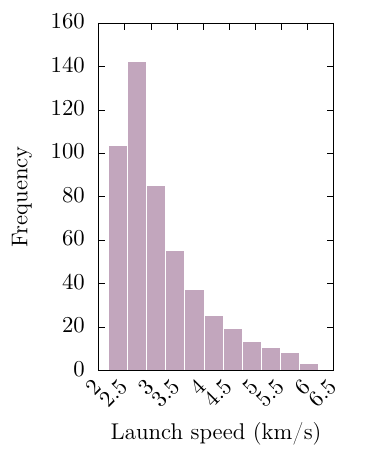} 
    \end{subfigure}
    \caption{ (Left) Initial velocity vectors of particles launched from an impact. Three sets of sites are considered, one in the North Hemisphere, one at the Equator and one in the South Hemisphere of the Moon. At each latitude we choose four launching sites at the lunar near-side, trailing-side, far-side and leading-side. From each site 500 particles were launched on a cone with a $45^{\circ}$ opening angle and random azimuthal angles (all the lengths were drawn equal for clarity). (Right) Histogram of launch speed distribution for lunar ejecta, showing 500 simulated speeds generated based on the power-law  distribution reported by \cite{Jiao2024}. The distribution spans from 2.38 $\mathrm{km/s}$ to 6.0 $\mathrm{km/s}$. } 
    \label{fig:launching}
\end{figure}

The initial speeds of the particles were drawn from the power-law distribution reported for escaping lunar fragments of tens of meters in size \citep{Jiao2024}, such as Kamo'oalewa and 2024 PT5, produced by a collider with a speed of 18 m/s and an impact angle of $45^\circ$.  
Hydrodynamic simulations of impact cratering  show that most ejecta are launched at angles between $30^\circ$ and $45^\circ$, with the latter being a reasonable representative value for these vertical impacts \citep{Melosh_1989,Jiao2024}. 
This angle maximizes the ejection efficiency while balancing the likelihood of particles achieving escape velocity. 
Therefore, in our simulations, at each launch location, particles were launched within a cone of $45^{\circ}$ aperture around the local vertical, with azimuthal angles chosen randomly (see Fig~\ref{fig:launching} (Left)). 
The random azimuthal angle assumes isotropic ejection, a reasonable assumption given the near-circular shapes of the impact craters. 

This distribution spans from 2.38 $\mathrm{km/s}$, the lunar escape velocity, to 6.0 $\mathrm{km/s}$, with higher speeds occurring progressively less frequently.
The velocity distribution is shown in Fig~\ref{fig:launching} (Right). 

For each launch site, 500 particles were launched with the velocities generated as previously described. 
This choice reflects an intermediate sampling size based on estimates of escaping fragments from major impact events; 
the Giordano Bruno crater impact, as modeled by \cite{Jiao2024}, could have generated 100–400 escaping fragments larger than 36 meters, while a larger impact event, such as the one that formed the Kepler crater, is estimated to have produced over 1,000 escaping fragments.
Other recent lunar craters younger than $\sim20$ million years are smaller than the Giordano Bruno crater, implying that fewer ejected fragments might be expected from them \citep{Mazrouei2019}. \\ 

Compared to the launching conditions used in \cite{Castro2023,Castro2025}, which focused on the longitudinal hemispheres of the Moon with sites restricted to the equatorial plane, the current study expands to more general scenarios. 
Here, particles are allowed to be launched out of the ecliptic plane and their initial launch speeds follow a more accurate power-law distribution, rather than the uniform sampling previously employed. \\ 

A total of 6,000 particles were launched. These TP’s were launched at a specific epoch $t_{0} =$ J2452996 (22 December 2003) and the initial conditions for the planets and the Moon were retrieved from the JPL Horizons web-service.
Previous studies \citep{Castro2025} have examined the sensitivity of lunar ejecta evolution to the choice of initial epoch, showing that the secular variation of Earth's orbit over the past megayear does not significantly affect the long-term dynamical evolution of the particles. Therefore, our choice of initial epoch is not expected to introduce substantial biases in the results.

\section{Numerical Results}
\label{sec:results}

We integrated the launched particles for a total time of 100,000 years. 
We detected collisions only with the Moon and the Earth; 
no collisions were detected with other bodies in the model (the Sun and the other planets).
Table~\ref{table:summary} summarizes the outcomes of 6,000 particles launched across the Northern, Equatorial, and Southern latitudes, divided further by their initial launching sites on the near, trailing, far, and leading sides of the Moon. 
Of the total particles, 0.9\% collided with the Moon, while 22.6\% collided with the Earth. 
The largest number of Earth impacts occurred for particles launched from the lunar trailing hemisphere, consistent across all latitudes; 
in contrast, particles launched from the lunar leading hemisphere exhibited the fewest Earth collisions. The Earth collision rates for equatorial and southern hemisphere launched particles are remarkably similar. 
The rates for the Northern hemisphere are slightly larger. The vast majority of the non-colliding particles ($\sim 63$\%) evolve into Aten or Apollo-like orbits as previously reported in \citet{Castro2023}; in the longer numerical integrations reported here, we also found that a small fraction ($\sim 11$\%) of the non-colliding particles also migrated into near-Venus space, suggesting that lunar ejecta could be mixed in with the larger population of NEOs. \\

Our overall Earth collision rate of 22.6\% aligns well with that of \cite{bG95} who reported that 20–25\% of lunar ejecta would collide with Earth over $\sim 10^{5}$-year timescales. (We note that \cite{Castro2023} reported Earth collision rates of $\sim 8$\% for equatorial launch sites, but over the shorter integration times of 5,000 years.)
The fraction of particles colliding with the Moon was small (0.9\%), consistent with previous studies, such as those by \cite{bG95} and \cite{Jiao2024}, showing that lunar ejecta that are launched with greater-than-escape speed have a low probability to be recaptured. 

\begin{table}[ht]
\begin{tabular}{l|ll|ll|ll}
         & \multicolumn{2}{c|}{North} & \multicolumn{2}{c|}{Equator} & \multicolumn{2}{c}{South} \\ \hline
Side     & Moon        & Earth        & Moon         & Earth         & Moon        & Earth       \\ \hline
Near     & 1           & 124          & 5            & 112           & 10          & 111         \\ \hline
Trailing & 9           & 149          & 3            & 137           & 6           & 139         \\ \hline
Far      & 2           & 120          & 5            & 113           & 5           & 108         \\ \hline
Leading  & 2           & 98           & 2            & 70            & 2           & 77          \\ \hline
         & 14          & 491          & 15           & 432           & 23          & 435        
\end{tabular}
\caption{Summary of collisional fates of the 6000 simulated particles after $10^{5}$ years. The total number of particles colliding with the Moon is 52 (0.9\%), and the particles colliding with the Earth is 1358 (22.6\%).} 
\label{table:summary}
\end{table}

\subsection{ Temporal distribution of collisions}

\begin{figure}[ht]
    \centering
    \includegraphics[width=0.4\textwidth]{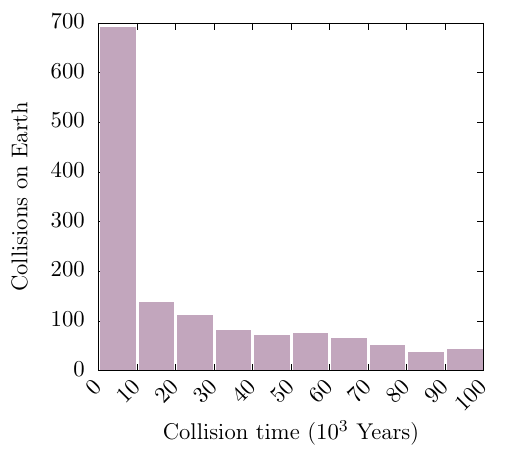}  
    \caption{Histogram of collision times for the simulated lunar ejecta impacting Earth. Most collisions occur within the first 10,000 years.}
    \label{fig:col_times}
\end{figure}

\begin{figure}[ht]
    \centering
    \includegraphics[width=0.75\textwidth]{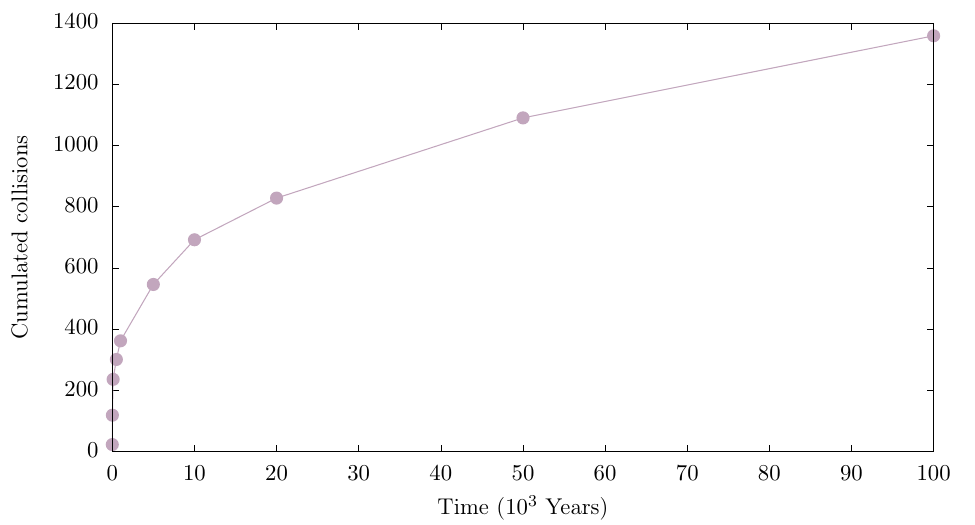}   
    \caption{Cumulative number of lunar ejecta collisions with Earth as a function of time. The curve shows a rapid increase of collisions during the first 10,000 years, followed by a slower more steady accumulation until the end of the simulation time span.}
    \label{fig:cumulated_1}
\end{figure}

\begin{figure}[ht]
    \centering
    \includegraphics[width=0.75\textwidth]{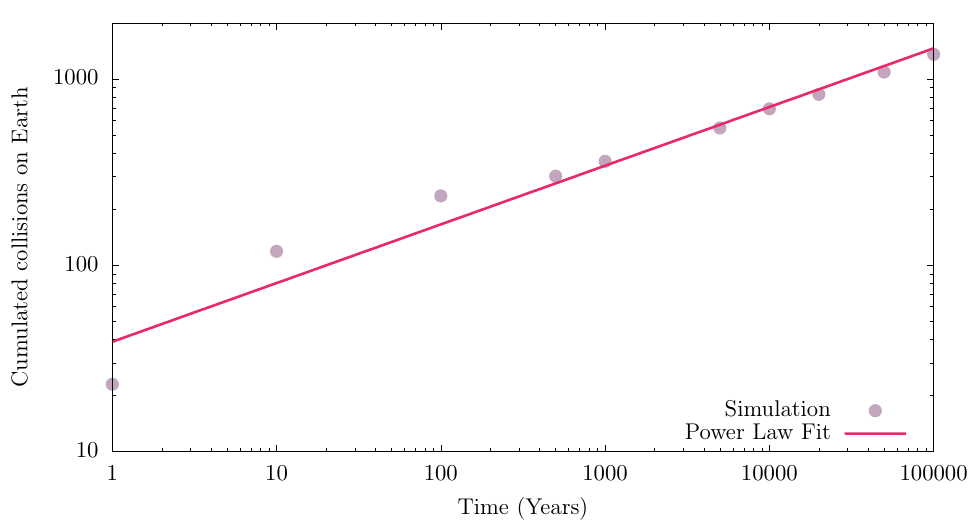}  
    \caption{Cumulative number of lunar ejecta collisions with Earth over time, on a log-log scale. This scale shows that the late behavior is consistent with a power-law (see Eq.~(\ref{eq:power_law})). }
    \label{fig:cumulated_2}
\end{figure}

The temporal evolution of lunar ejecta that collided with Earth was analyzed using the distribution of collision times and cumulative collision counts over the 100,000-year simulation period. 
These results are shown in Figures ~\ref{fig:col_times}. \ref{fig:cumulated_1}, \ref{fig:cumulated_2}. 

Figure~\ref{fig:col_times} displays the distribution of collision times for ejecta particles. 
The majority of collisions with Earth occur within the first 10,000 years following ejection, representing a sharp initial peak in collision frequency. 
Beyond that time span, the collision frequency decreases significantly, but with a long tail extending over the remaining 90,000 years. 
This suggests that a small fraction of ejecta maintain Earth-crossing orbits over extended timescales ($\sim 3\%$ of the collided particles were delivered in the final 10,000 years). This is consistent with the trasnfer times derived from cosmic ray exposure of known lunar meteorites, ranging from few thousands of years to few millions of years \citep{Joy2023,Nishiizumi2009,Lorenzetti2005}.  \\

Figure~\ref{fig:cumulated_1} shows the cumulative number of collisions as a function of time. 
The curve steeply rises during the initial 10,000 years, reflecting the dominance of early collisions. 
After this, the curve flattens, indicating a reduced collision rate over longer timescales. 
This coincides with the median of the distribution, since half of the recorded collisions occur in the first $\sim 9,500$ years, suggesting that this is a representative timescale for lunar impact generated flux at Earth. 

More insights can be gained by portraying the cumulative number of collisions in a log-log scale, as shown in Fig~\ref{fig:cumulated_2}. 
The linear trend observed at later times (after about 10,000 years) follows a power-law relationship. 
Based on the line of best fit, the number of acccumulated collisions $C(t)$ is 
\begin{equation}
    C(t) = (38.9 \, \mathrm{collisions}) \, t^{0.315}, \qquad \hbox{for\, $t$ > 10,000\, years}
    \label{eq:power_law}
\end{equation}
where $t$ is measured in years, and the constant factor is specific to our total number of 6000 simulated particles. 
A similar trend is observed for each of the individual  sites (see Figure~\ref{fig:times_sites} in Appendix~\ref{sec:sample:appendix}).  
Variations in the power-law exponent range from $0.284$ (trailing side) to $0.35$ (leading side). 
Using the previous kind of law, we can estimate that the number of surviving particles from the far side after 10 Myr is 7\%. This is smaller by about a factor of 2 than the value of 16\% found by Jiao et. al. (2024), indicating that this power law overestimates the number of colliders over very long times. 
Based on the different exponents found, the maximum range of validity of this kind of power law is in the range of 3-20 Myr. \\

\subsection{ Impact velocity distributions}

\begin{figure}[ht]
    \centering
    \begin{subfigure}[b]{0.45\linewidth}
        \centering
        \includegraphics[width=0.82\textwidth]{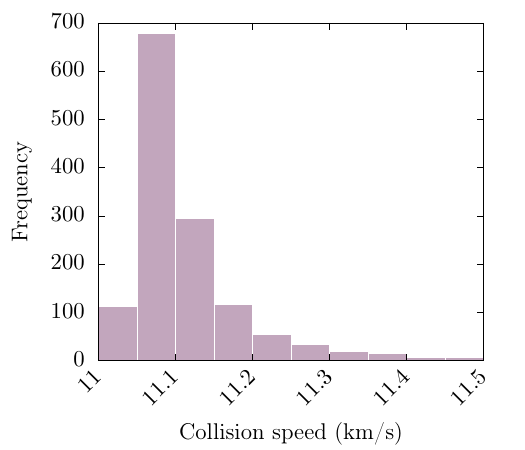}   
    \end{subfigure}
    \begin{subfigure}[b]{0.45\linewidth}
        \centering
        \includegraphics[width=0.98\linewidth]{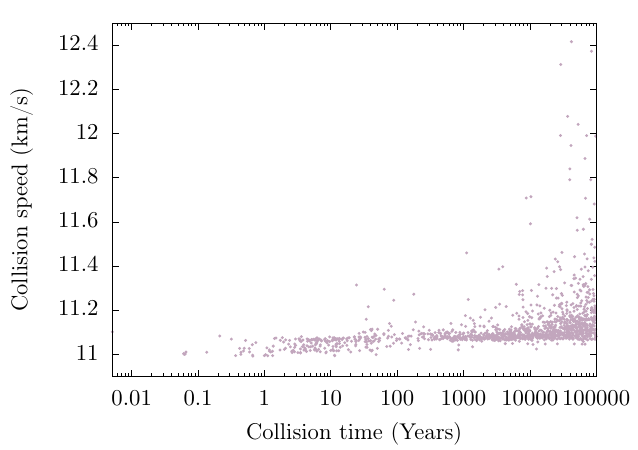}
    \end{subfigure}
    \caption{ (Left) Distribution of impact speeds with Earth's atmosphere for the simulated lunar ejecta. The distribution peaks below these values. (Right) Scatter graph of time to collision (in log scale) and the speed at atmospheric entry for lunar eject. The plot indicates that particles taking longer to collide with Earth tend to have higher speeds at atmospheric entry, while those colliding shortly after ejection exhibit lower speeds. } 
    \label{fig:speeds}
\end{figure}

The speeds of lunar ejecta upon collision with Earth's atmosphere are shown in Fig.~\ref{fig:speeds} (Left). 
The velocities range from 11.0 to 13.1 $\mathrm{km/s}$, with a clear peak at 11.05-11.10 $\mathrm{km/s}$. 
One of the main factors determining the value of the speed is the time to impact. 
Figure~\ref{fig:speeds} (Right) shows that particles that collide shortly after ejection exhibit speeds near the lower end of the range, while particles that take longer to hit the Earth tend to have higher velocities (all cases of Earth collision speed exceeding 12.0 $\mathrm{km/s}$ take longer than 20,000 years to collide). 
This behavior aligns with the dynamics of orbital interactions. 
Particles with shorter collision times follow nearly direct trajectories, while particles on extended dynamical paths experience more complex orbital evolution, including interactions with Earth's (or other planet's) gravity well, leading to increased entry velocities. 

\subsection{ Geographic and temporal patterns of impacts on Earth} 
\label{subsec:geo_time}

\begin{figure}[ht]
    \centering
    \begin{subfigure}[b]{0.45\linewidth}
        \centering
        \includegraphics[width=0.88\textwidth]{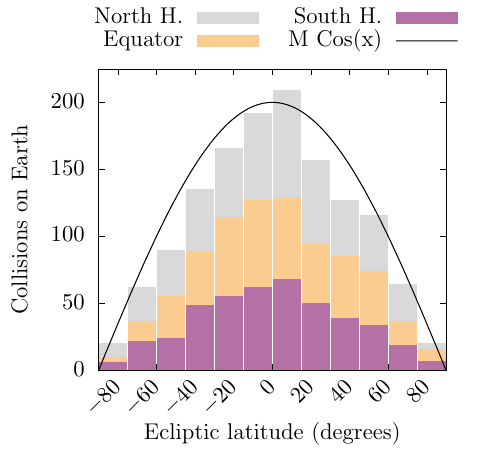}   
    \end{subfigure}
    \begin{subfigure}[b]{0.45\linewidth}
        \centering
        \includegraphics[width=0.95\textwidth]{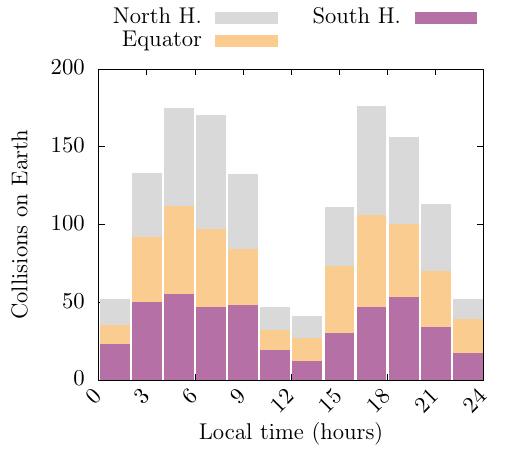} 
    \end{subfigure}
    \caption{ (Left) Distribution of the ecliptic latitude of lunar ejecta collisions with Earth's atmosphere. The distribution clearly peaks near the equator. The contributions of colliders launched from the Moon's Northern and  Southern hemispheres and and from the lunar Equator are shown in \textit{gray}, \textit{violet} and \textit{yellow} respectively. For comparison, the solid black line shows a cosine function representing a uniform area distribution. (Right) Distribution of local collision times for lunar ejecta impacting Earth. } 
    \label{fig:latitudes_times}
\end{figure}

A significant fraction of impacts occur near the equator, with the frequency decreasing toward higher latitudes. 
Figure \ref{fig:latitudes_times} (Left) presents the ecliptic latitude distribution of collisions with Earth. While a uniform surface distribution would correspond to cosine function, our results show a thinner distribution, slightly skewed toward mid-latitudes in both hemispheres. 
This can be confirmed by calculating the flux of colliders per unit area. 
The surface area of a band between latitudes $\lambda_{1}$ and $\lambda_{2}$ is calculated using
\begin{equation}
    A_{\mathrm{band}} = 2 \pi R_{\mathrm{Earth}}^{2} \left( \sin{\lambda_{2}} - \sin{\lambda_{1}} \right).   
\end{equation}
With this, we calculate that the ratio of the flux at the polar and the equatorial bands is 0.76. 
These findings can be compared with those of \cite{LeFeuvre2008}, who found a ratio of 0.90. 
The previous work is based on the known models of the NEO population, therefore, our results suggest that lunar ejecta orbits tend to have lower relative inclinations than the general population of NEOs.However, this distribution may evolve over time; our results indicate that at early times, the Earth impactors are more concentrated at lower ecliptic latitudes, whereas at later times, their ecliptic latitude distribution flattens, and their impact flux per unit area becomes more uniform. 

In our simulations, we do not explicitly account for Earth's rotation. As a result, while we cannot determine the longitude of an impactor at the top of Earth's atmosphere at the moment of collision, we can still determine the local time of impact by examining the relative position of the particle with respect to Earth and the Sun; that is, noon (12:00) corresponds to when the Sun is directly overhead and midnight (0:00 or 24:00) corresponds to when the location is on the opposite side of Earth from the Sun.  

From our simulation results, we estimate that AM colliders represent 52.2\% of the total, and PM colliders the remaining 47.8\%. 
The distribution of local times is shown in Fig~\ref{fig:latitudes_times} (Right), indicating a clear increase in impacts close to the 06:00 and 18:00, with the number of events decreasing significantly around noon and midnight.  
\cite{Gallant2009} found that local time distribution was a strong function of the speed at impact, identifying large asymmetries depending on the speed ranges of the sample (their impact speed range was 11 -- 30 $\mathrm{km/s}$). 
In our simulations, the range of speeds at Earth impact of the simulated lunar ejecta is very narrow (see Fig.~\ref{fig:speeds}), and we do not find such asymmetries in the AM/PM distribution. 

The diurnal peak coincides with the leading side of the Earth facing incoming particles, and thus crossing more particles and increasing the number of collisions. 
The small range of speeds at impact, and its values being so close to the Earth's escape velocity indicates that the colliding particles have very Earth-like orbits and may also be able to reach the Earth from its trailing side, explaining the nocturnal peak at 18:00. \\ 

\section{Discussion} 
\label{sec:Discussion}

\subsection{Biological and geological implications}

The impact of extraterrestrial objects on Earth has played a significant role in shaping our planet's biological and geological history \citep{Zahnle2006}. 
Lunar ejecta, generated by meteoroidal impacts on the Moon, contribute to the transport of materials between solar system bodies and have been a subject of study to understand these processes. 
However, our analysis of lunar ejecta flux suggests that these particles do not pose a substantial threat to Earth's habitability or geological stability under current conditions. 
To further contextualize the potential impact of lunar ejecta, we can compare them with more recent events like the Chelyabinsk meteor. 
The Chelyabinsk event in 2013 involved a near-Earth asteroid approximately 20 meters in diameter entering Earth's atmosphere at about 19 $\mathrm{km/s}$. 
This event caused minor building damage and injuries due to broken windows \citep{Popova2013}. 
While the sizes of the lunar ejecta considered in our simulations may be similar to that of the Chelyabinsk meteorite, their velocities upon collision with Earth's atmosphere are smaller, 11.1 -- 13.1 $\mathrm{km/s}$. 
Such impactors would release half as much energy as the Chelyabinsk meteorite and would typically leave no crater, although fragments could reach the surface \citep{Collins2017}. 
The flux of these falling fragments would be larger at the equatorial latitudes as discussed in \textsection\ref{subsec:geo_time}. 

\subsection{Contribution to the population of NEOs}

The existence of lunar-originated NEO has been discussed in recent years. 
\cite{bS21} suggested that the NEA Kamo'oalewa could be a fragment of the Moon, based on its reflectance spectrum and Earth-like orbit. 
Additional evidence based on orbital evolution and hydrodynamic impact simulations supported this possibility \citep{Castro2023,Jiao2024}. 
More recently, \cite{Kareta2025} presented evidence that the NEA 2024 PT (originally referred to as a 'Mini-Moon' \citep{delaFuenteMarcos:2024}) had spectral properties very similar to known lunar materials. 
That study also identified further candidates of lunar-derived objects. 
Current NEO population models primarily focus on objects originating from the asteroid belt and do not account for the possible existence of a lunar contribution to the NEO population \citep{Nesvorny2024}, though this population size would be small. 

\begin{figure}[ht]
    \centering
    \begin{subfigure}[b]{0.475\linewidth}
        \centering
        \includegraphics[width=0.95\textwidth]{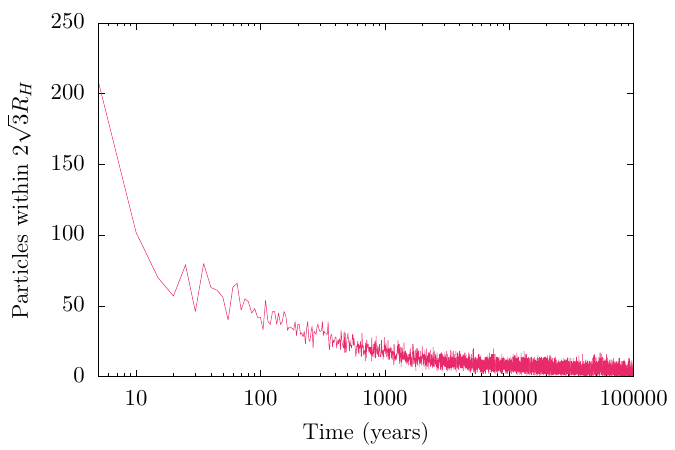}   
    \end{subfigure}
    \begin{subfigure}[b]{0.475\linewidth}
        \centering
        \includegraphics[width=0.95\textwidth]{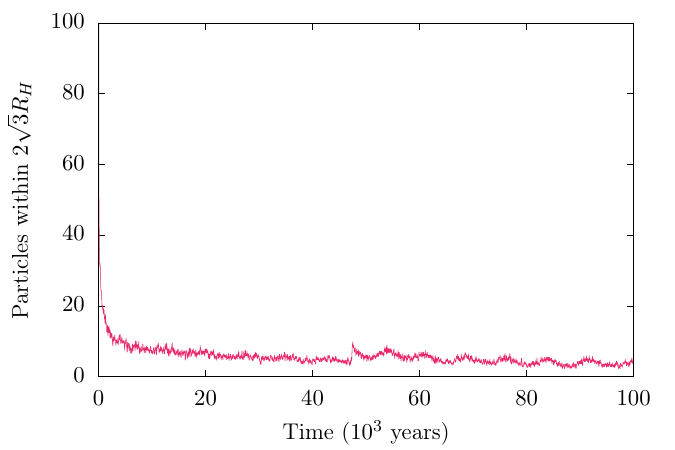} 
    \end{subfigure}
    \caption{ (Left) Number of particles (from an initial 6000 particles set) inside the sphere of radius $2\sqrt{3} R_{H}$ around the Earth as a function of time (in logarithmic scale). The number of particles in this region decreases rapidly and reaches a small asymptotic value at long times. (Right) Average number of particles (in a 100 years window) inside a sphere of radius $2\sqrt{3} R_{H}$ around the Earth as a function of time. The average asymptotic value is $\sim 4$ particles.} 
    \label{fig:coorbitals}
\end{figure}

Our results suggest that a fraction of lunar ejecta may persist in the vicinity of the Earth for extended periods, contributing to the observed diversity of small bodies in near-Earth space. 
Figure~\ref{fig:coorbitals} shows the number of particles within the stability zone of the Earth, inside a sphere of radius $2\sqrt{3} R_{H}$, where $R_{H}$ is Earth's Hill's radius \citep{Gladman1993}. 
A similar behavior is observed if we instead consider the particles inside the Hill sphere of the Earth, with the asymptotic average value decreasing to $\sim 0.1$ particles in the region. 
As our sampling frequency is one output every 5 years, the mean waiting time to observe an object inside Earth's Hill sphere within our simulated sample is 50 years. 
Meanwhile, there exist 3 craters larger than 10 km younger than 20 Myr \citep[e.g.][]{Mazrouei2019}  , our estimation for the maximum range of validity for our results. 
With our model assumptions, we estimate that $\sim 350$ lunar ejecta fragments (each tens of meters in size) have impacted Earth in the past 20 million years, with the rest $\sim 1,150$ migrating elsewhere in near-Earth space and $\sim1$ of them passing inside Earth's Hill sphere every 200 years. This impact rate is about 3.5\% of the impact rate of similar-sized NEOs \citep{NASA_Asteroid_Facts}, such impacts would be concentrated within a few thousand years following their ejection and would favor low ecliptic latitudes.

\subsection{Future Work}

Our results are based on lunar ejecta model that assumes vertical impacts, which is a widely used simplification in impact cratering studies. However, natural impacts are predominantly oblique, which significantly affects the distribution of escaping ejecta. Studies using 3D hydrodynamic simulations  \citep{Artemieva2019,Luo2022} have shown that oblique impacts introduce substantial differences with the launching conditions we considered. For instance, these impacts can produce ejection angles lower than $30^{\circ}$, instead of the assumed $45^{\circ}$. The direction of the ejecta would be highly anisotropic, with a large fraction of the material going in the impact direction. These factors suggest that our estimated Earth impact fluxes may be conservative, as oblique impacts may produce more Earth-crossing ejecta than vertical ones. While incorporating full 3D oblique impact models is beyond the scope of this study, future work should explore a broader range of ejection conditions. \\

Our study has focused on the current dynamics of lunar ejecta.
However, the Moon's geological history indicates that most of the significant impact events occurred during ancient times, particularly before the Copernican period (which began approximately 1.1 billion years ago). 
During the pre-Nectarian and Nectarian periods, the Moon experienced intense bombardment, leading to the formation of numerous large basins and craters \citep{Wilhelms1987}. 
It would be of interest to investigate the orbital dynamics of lunar ejecta during these ancient periods, taking into account the Moon's historical orbital configurations \citep[e.g.][]{Bills1999}. 
We would expect that a closer Moon would likely result in a higher fraction of lunar ejecta landing on Earth and a lower fraction migrating to heliocentric space; moreover, due to the more intense bombardment, we would expect an overall higher flux of lunar ejecta.
Such a study could enhance our understanding of the early Earth-Moon system and the potential exchange of materials between these two bodies during critical periods of planetary development.

\section{Conclusions}

\begin{itemize}
    \item Following an impact event on the Moon, half of all lunar ejecta collisions with Earth occur in a timespan of $\sim$10,000 years. The cumulative number of collisions at later times follows a power-law distribution, $C(t) \propto   t^{0.315}$. 
    \item Approximately 22.6\% of simulated lunar ejecta collide with Earth over 100,000 years. Particles launched from the trailing side of the Moon have the highest Earth collision rates, while those launched from the leading side have the lowest rate. 

    \item Impact velocities at Earth of lunar ejecta are in the range 11.0–13.1 $\mathrm{km/s}$, and the impact locations are geographically concentrated near equatorial latitudes, with a polar-to-equatorial flux ratio of 0.76. PM and AM impactors are mostly symmetrically distributed with similar total counts and peaks near 6 PM/AM. 
    \item Lunar ejecta impacting Earth are, in general, slower and more concentrated toward equatorial latitudes than those from the general NEO population. 
\end{itemize}

 \section{Data Availability}

Binary files with the initial conditions used for the simulations presented in this paper will be available at a publicly accessible permanent repository after the paper is accepted. 

\section{Acknowledgments}
The results reported herein benefited from collaborations and/or information exchange within the program ``Alien Earths'' (supported by the National Aeronautics and Space Administration under Agreement No. 80NSSC21K0593) for NASA’s Nexus for Exoplanet System Science (NExSS) research coordination network sponsored by NASA’s Science Mission Directorate. A.R. acknowledges support by the Air Force Office of Scientific Research (AFOSR) under Grant No. FA9550-24-1-0194.

\printcredits

\bibliographystyle{cas-model2-names}

\bibliography{cas-refs}


\appendix
\section{Appendix}
\label{sec:sample:appendix} 

\begin{figure}[ht]
    \centering
    \includegraphics[width=1.0\linewidth]{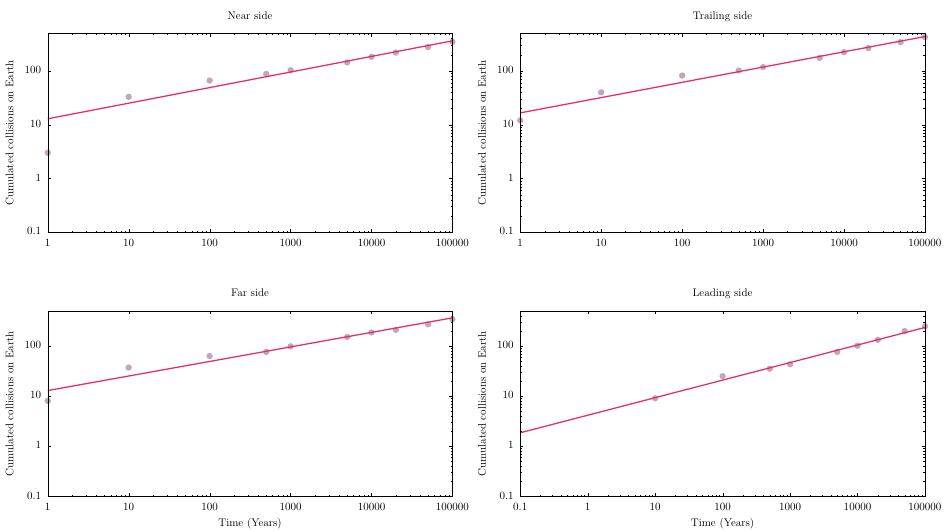}
    \caption{Cumulative number of lunar ejecta collisions with Earth over time, on a log-log scale for particles launched from each lunar side. The power-law fit is shown for each of them.}
    \label{fig:times_sites}
\end{figure}

\end{document}